\documentclass{article}
\usepackage{epsfig}
\newcommand{\bfr}{\begin{flushright}}
\newcommand{\efr}{\end{flushright}}
 
\begin{document}
\title{Euclidean wormhole solutions of Einstein--Yang--Mills theory in
diverse dimensions
}
\author{Katsuhiko Yoshida and Satoru Hirenzaki\\
Department of Physics,
Tokyo Metropolitan University,\\ Setagaya, Tokyo 158, Japan\\
Kiyoshi Shiraishi\\
Institute for Nuclear Study, University of Tokyo,\\ Midori-cho, Tanashi,
Tokyo 188, Japan
}
\date{Phys. Rev. \textbf{D42}, 1973Ð1981 (1990)
}
\maketitle
\begin{abstract}
We solve the Euclidean Einstein equations with non-Abelian gauge
fields of sufficiently large symmetry in various dimensions. In
higher-dimensional spaces, we find the solutions which are similar to
so-called scalar wormholes. In four-dimensional space-time, we find
singular wormhole solutions with infinite Euclidean action. Wormhole
solutions in the three-dimensional Einstein--Yang--Mills theory with a
Chern--Simons term are also constructed.
\end{abstract}

\section{INTRODUCTION}
Various attempts to quantize gravity have been made
by many authors. The path-integral quantization of gravity \cite{1}
has been a very popular approach to quantization in
the past decade.

A few years ago, a very attractive mechanism to determine
the ``constants'' in nature in the framework of quantum
gravity was suggested \cite{2}. The mechanism is based on
the appearance of a wormhole configuration in the Euclidean
path integral over the distinct topology of spacetime.
In particular, the wormhole and baby universes
connected by the wormholes have been invoked to explain
the vanishing cosmological constants \cite{2,3}.

Many problems were pointed out soon after this suggestion,
however. For example, the treatment of contributions
of large wormholes \cite{4}, the efficiency of the dilute gas
approximation \cite{5}, and the phase of the result of the
path integral \cite{6} have been discussed. The validity of Euclidean
quantum gravity itself has also been investigated.

Nevertheless we must equip our ``theoretical arena'' by
studying the wormhole solutions to Einstein equations
coupled to possible matter fields, until the refined formulation
of quantum gravity appears. It is suggested that
the correct procedure is to sum only over stationary
points of the Euclidean action \cite{7}. Thus, we would like to
study the various types of wormhole configurations.

We have already known many kinds of wormhole solutions.
The following matter fields which support the
``throat'' of the wormhole were adopted: axion fields \cite{8},
scalar fields with and without spontaneous breaking of
global $U(1)$ symmetry \cite{9,10,11,12,13,14,15}, and $SU(2)$
Yang--Mills fields \cite{16,17}. Higher-dimensional wormhole
solutions were also considered \cite{18}, and the higher-derivative
correction to the Einstein--Hilbert action was investigated by several
authors \cite{19}.

Among them, ``gauge field wormholes'' in Refs.~\cite{16} and
\cite{17} have very attractive features. Non-Abelian gauge
fields are believed to be fundamental entities in unified
theories including superstring(-inspired) models \cite{20}.
Therefore gauge field wormholes are considered as inevitable
objects in the Euclidean formulation. Recently, an
interesting class of wormhole solutions has been constructed
for the $SU(N)$ Yang--Mills system \cite{21}. These
wormholes have a static structure of a ``magnetic'' gauge
field configuration as the $SU(2)$ Yang--Mills wormhole
offered in Ref.~\cite{16}. On the other hand, in the same $SU(2)$
case, the solution to Yang--Mills equations which exhibit
periodic ``motion'' in Euclidean time was obtained by the
author of Ref.~\cite{21} recently. Thus we expect the existence
of the wormhole solutions which involve two more
dynamical variables, in the case where the gauge system
has a higher symmetry than $SU(2)$.

In this paper we construct wormhole solutions in the
viewpoint of dimensional reduction of gauge fields.
Non-Abelian gauge fields can be symmetrically reduced
to effective self-interacting scalar fields by the method of
coset-space dimensional reduction \cite{22,23,24}. We consider a
$(1+d)$-dimensional space-time. In our case, the whole
``space,'' namely, $S^d$, is the coset space. We will try to
make use of the effective scalar fields to construct the
wormhole solutions similar to scalar wormholes.
We shall not discuss the stability of the wormhole solutions.
Note, however, that classically unstable solutions
can largely contribute to the path integral in some special
circumstances.

This paper is organized as follows. In Sec.~II, we briefly
review the coset-space reduction of Yang--Mills fields,
which is necessary to construct wormhole solutions. We
then consider first the higher-dimensional cases (Sec.~III).
In Sec.~IV we investigate the wormholes in 1+3 dimensions.
The form of the solution is very different from the
solution already known. In Sec. V we present the
wormhole solution in $1+2$ dimensions. The Chern--Simons term in the
action is essential to construct the solution. Finally, Sec.~VI is
devoted to the conclusion.

\section{REDUCTION OF GAUGE FIELDS}
As is well known, a non-Abelian gauge theory symmetric
with respect to the canonical action of $G$ on the
manifold $M \times G/H$ can be reduced to a gauge theory in
M that includes scalar fields coupled to the gauge
field \cite{22,23,24}. Many people have studied the reduction
schemes of the Yang--Mills field in the context of Kaluza-Klein
compactification\cite{25}. The scalar potential in the large number of
reduction schemes is the Higgs-type potential.

To apply the technique to our system, we use the $d$-dimensional
space as a coset space. Now we take the ansatz
for the structure of Euclidean space-time. We assume
that the wormhole solution has spherical symmetry.
The metric can be written as $ds^2=dr^2+a^2(r)d\Omega^2(S^d)$,
where $d\Omega^2(S^d)$ is the line element of the $d$-sphere with
unit radius. We consider the symmetric reduction of
gauge fields with symmetry group $K$.

The sphere $S^d$ is realized in the form of the symmetric
spaces $SO(d+1)/SO(d)$. If the isotropy group $H$ is simple
such as $SO(d)$, then for the simplest embeddings and
rank $G\le \mbox{rank~} K$ we always have $H\subseteq G\subseteq K$, and
vice versa. Moreover, if the embedding $H\subseteq K$ exists for which
the homomorphism $\{\tau: H\rightarrow K\}$ can be extended to a
homomorphism of $G$ to $K$, the minimum value of the
effective scalar potential is zero \cite{23}. Thus in our case, for a
sufficiently large gauge group $K\supseteq SO(d+1)$, the
minimum value of the potential is zero.

In a typical case, the potential of the reduced theory is
of the form \cite{23,24}
\begin{equation}
V(\Phi)\propto  (|\Phi|^2-\lambda^2)^2\,,
\label{2.1}
\end{equation}
where the complex variable $\Phi$ comes from the mapping
$S^d$ to the gauge group $K$. $\Phi(\tau)$ behaves as a scalar field.
The potential reaches the minimum value $V=0$ at $|\Phi|^2=\lambda^2$.
Here $\Phi$ has several complex components in
general.

The potential comes from ${\rm Tr~} F_{ij} F^{ij}$ ($i,j=1,\dots,
d$) in the original Yang--Mills theory. On the other hand, the kinetic
term for the effective scalar comes from ${\rm Tr~} F_{0i} F^{0i}$.

The potential depends on $\Phi$ only through $|\Phi|$. Then
we obtain a ``wine-bottle''-type potential, and the minima
of the potential are mutually connected by the gauge rotation
that generates the bottom of this potential.

If we apply the reduction scheme to the construction of
the wormhole solutions, we have only to consider the
simplest situation. The authors of Ref.~\cite{15} showed that
the field equations in non-Abelian scalar systems are reduced
to be essentially the same as the equations in the
$U(1)$-symmetric case described in Refs.~\cite{9,10,11,12,13}.
Therefore, we will consider here $U(1)$-symmetric potential of the
wine-bottle type, even if the reduced theory has more
symmetries; in that case we simply set the effective field
to zero except one complex effective scalar. This ``ansatz''
is consistent with the field equations in the case
with the effective potential of the type (\ref{2.1}). We shall restrict
ourselves to this class of the effective potential only.

In general, we can rescale the effective scalar field and
express the reduced Yang--Mills term as
\begin{equation}
{\rm Tr~} F^{MN} F_{MN}= 4d\alpha |\dot{\Phi}|^2/a^2+
2d(d-1)\beta(|\Phi|^2-1)^2/a^4\,,
\label{2.2}
\end{equation}
where the numerical coefficients $\alpha$ and $\beta$ are expected to
be $\sim 1$, and $M$ and $N$ run over $0, 1, \dots , d$. The complex
variable $\Phi$ plays a role of a scalar field. Here we set the
zeroth component of the gauge field to zero. This is
physically justified by a gauge choice. Note that the reduced
Lagrangian for the effective scalars depends on the
radius of the sphere, $a$.

An example of the representation of $SU(N)$ gauge fields
on $S^d$ is shown in the Appendix. Since the mapping is
not always expressed in so simple a form for arbitrary $d$
and $N$, we show the simple cases where $N\ge d+1$.

In the next section we will apply the reduction (\ref{2.2}) to
the dynamics of Einstein--Yang--Mills system, and try to
find solutions of the wormhole type.

\section{HIGHER-DIMENSIONAL CASES}
We start from a coupled system of gravity and gauge
fields in the $(1+d)$-dimensional space-time. In this section
we consider the cases with $d\ge 4$. The action is
\begin{equation}
S=\int d^{d+1}x\sqrt{-g}\left[-\frac{1}{2\kappa^2}R+\frac{1}{4e^2}
{\rm Tr~}F^{MN} F_{MN}\right]+(\mbox{surface terms})\,,
\label{3.1}
\end{equation}
where $R$ is the scalar curvature and $F_{MN}$ is the field
strength of the non-Abelian gauge fields. The gauge symmetry
is assumed to be large enough to have an
$SO(d+1)$ group as a subgroup.

The field equations can be derived from the above action.
If we assume the metric $ds^2=d\tau^2+a^2(\tau)d\Omega^2(S^d)$,
and adopt the symmetric reduction of the form (\ref{2.2}), the
equations for $\Phi$ and $a$ are
\begin{eqnarray}
& &\ddot{\Phi}+(d-2)\frac{\dot{a}}{a}\dot{\Phi}=
\frac{d-1}{a^2}\frac{\beta}{\alpha}(|\Phi|^2-1)\Phi\,,\label{3.2a}
\\
& &\dot{a}^2=1+\frac{1}{d-1}\frac{\kappa^2}{e^2}\left[
2\alpha\frac{|\dot{\Phi}|^2}{a^2}-(d-1)\beta
\frac{(|\Phi|^2-1)^2}{a^4}
\right]\,,
\label{3.2b}
\end{eqnarray}
where an overdot denotes a derivative with respect to the
Euclidean time $\tau$. Equations (\ref{3.2a}) and (\ref{3.2b}) have been
derived from the Yang--Mills equation and the time-time
component of the Einstein equations.

When we write the complex scalar as $\Phi=f e^{i\psi}$, $\psi$ is a
cyclic variable in terms of the classical mechanics. The
equation for $\psi$ is integrable, and this variable is associated
with a conserved quantity.

Since we treat ``effective'' scalar systems and we are interested
in the construction of the solution in this paper,
we would like to sidestep the complete discussion on the
conserved charges and Euclidean formulation. Avoiding
the complicated issue about the treatment of conserved
charges, we take a conventional view \cite{11,13,14} of the integrable
equation.

Using the variables $f$ and $\psi$, one can read the following
equation from the imaginary part of the equation of
motion for $\Phi$:
\begin{equation}
(a^{d-2}f^2\dot{\psi})^\cdot=0\,. 
\label{3.3}
\end{equation}

After ``Euclideanization'', the normalized vector
directed in the time coordinate is $i\partial/\partial\tau$; so the
normalized derivative of the effective scalar field at the throat
takes the real value \cite{11,13,14}.  Thus we set
\begin{equation}
a^{d-2}f^2\dot{\psi}=in\,,
\label{3.4}
\end{equation}
where $n$ is a real integration constant. Note that the
``charge'' $n$ is not the electric or magnetic charge. In our
case, the integration constant corresponds to an ``external''
electric field. Thus we may expect particle creation
and quantum instability. We do not consider this an interesting
subject here, but we will report on it elsewhere.

Substituting new variables $f$ and (\ref{3.4}), the equations of
motion (\ref{3.2a}) and (\ref{3.2b}) become
\begin{eqnarray}
& &\ddot{f}+(d-2)\frac{\dot{a}}{a}\dot{f}=
\frac{\beta(d-1)}{\alpha a^2}f(f^2-1)-\frac{n^2}{f^3a^{2d -4}}\,,
\label{3.5a}\\
& &\dot{a}^2=1-\frac{\kappa^2}{e^2(d-1)}\left[
\frac{(d-1)\beta}{a^2}(f^2-1)^2+\frac{2\alpha n^2}{f^2a^{2d-4}}
-2\alpha\dot{f}^2
\right]\,.
\label{3.5b}
\end{eqnarray}

The ``centrifugal'' potential term which is proportional
to the square of the constant ``charge'' or ``angular
momentum'' holds the throat of the wormhole so it does
not shrink to a zero radius. The sign of the term relative
to the other terms is crucial for the existence of the
wormhole solutions.

Another important observation is the competition of
the centrifugal term and the original potential term. For
instance, the first term on the right-hand side of (\ref{3.5a}) is
proportional to $a^{-2}$, while the second term is proportional
to $a^{-(2d-4)}$. The behavior of the wormhole geometry
shows that the scale factor a approaches infinity when $\tau$
goes to infinity while the scale factor reaches a finite
value when $\tau=0$. If $d\ge 4$, the (attractive) ``centrifugal''
force dominates around $\tau\simeq 0$ and the potential force
dominates at infinity. This feature is analogous to that of
the usual scalar wormholes \cite{10}. On the other hand, if
$d\le 3$, $f$ and $a$ show quite different types of behaviors.
These subjects will be investigated in Secs.~V and VI.

The rescalings
\begin{equation}
\tau=KX\quad\mbox{with}\quad a=KA\quad\mbox{with} K=
\frac{1}{d-1}\frac{\kappa}{e}\frac{\alpha}{\sqrt{\beta}}
\label{3.6}
\end{equation}
lead to the field equations
\begin{eqnarray}
& &{f''}+(d-2)\frac{A'}{A}{f'}=
\frac{1}{L^2A^2}f(f^2-1)-\frac{2qL^2}{f^3A^{2d -4}}\,,
\label{3.7a}\\
& &\dot{A}^2=1-\left[
\frac{1}{4L^4A^2}(f^2-1)^2+\frac{q}{f^2A^{2d-4}}
-\frac{{f'}^2}{2L^2A^2}
\right]\,,
\label{3.7b}
\end{eqnarray}
with a prime denoting an $X$ derivative and
\begin{equation}
q=\frac{(d-1)\beta n^2}{2\alpha K^{2d-6}}\,,\quad 
L^2=\frac{\alpha}{(d-1)\beta}\,.
\label{3.8}
\end{equation}

The equations are very similar to those in Ref.~\cite{10} except
for the dependence of the terms on the scale factor $A$.

We have solved these coupled equations by using the
computer code named COLSYS \cite{27}. The resulting solutions
are shown in Figs.~1(a) and 1(b) for $d=4$, Figs.~2(a) and
2(b) for $d=5$, and Figs.~3(a) and 3(b) for $d=9$. All the
calculations in this section have been performed in the
case with $\alpha=\beta=1$.

\begin{figure}[ht]
\begin{center}
\includegraphics[width=5cm]{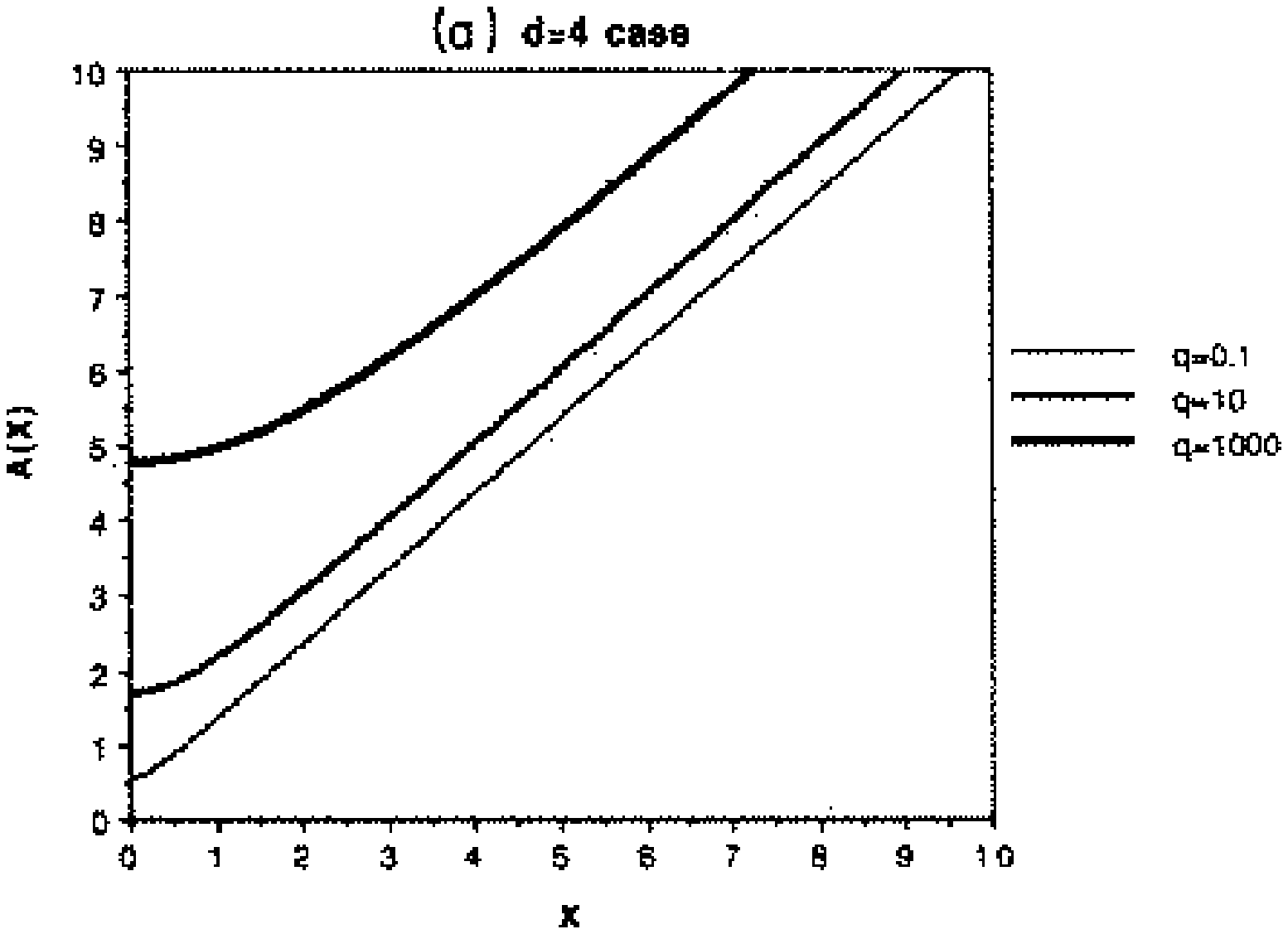}
\includegraphics[width=5cm]{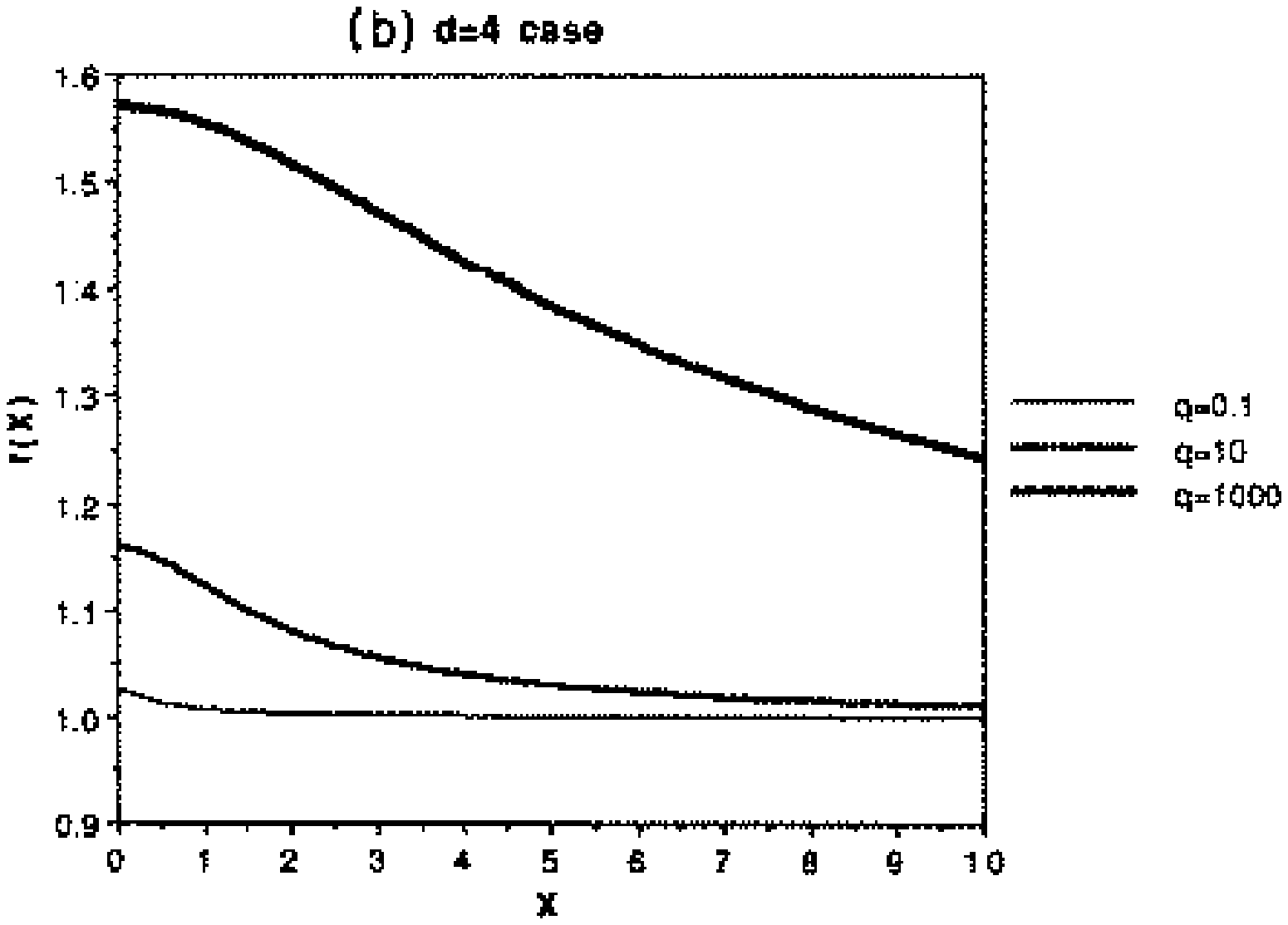}\\
\centering{(a)\hspace{5cm}(b)}
\caption{(a) $A(X)$ as a function of $X$ in the case where the
space dimension is four. The lines correspond to the parameter
$q =1000, 10,$ and $0.1$ in order of boldness. (b) $f(X)$ as a function
of $X$ in the case where the space dimension is four. The
definitions of lines are the same as in (a).}
\label{f1}
\end{center}
\end{figure}
\begin{figure}[ht]
\begin{center}
\includegraphics[width=5cm]{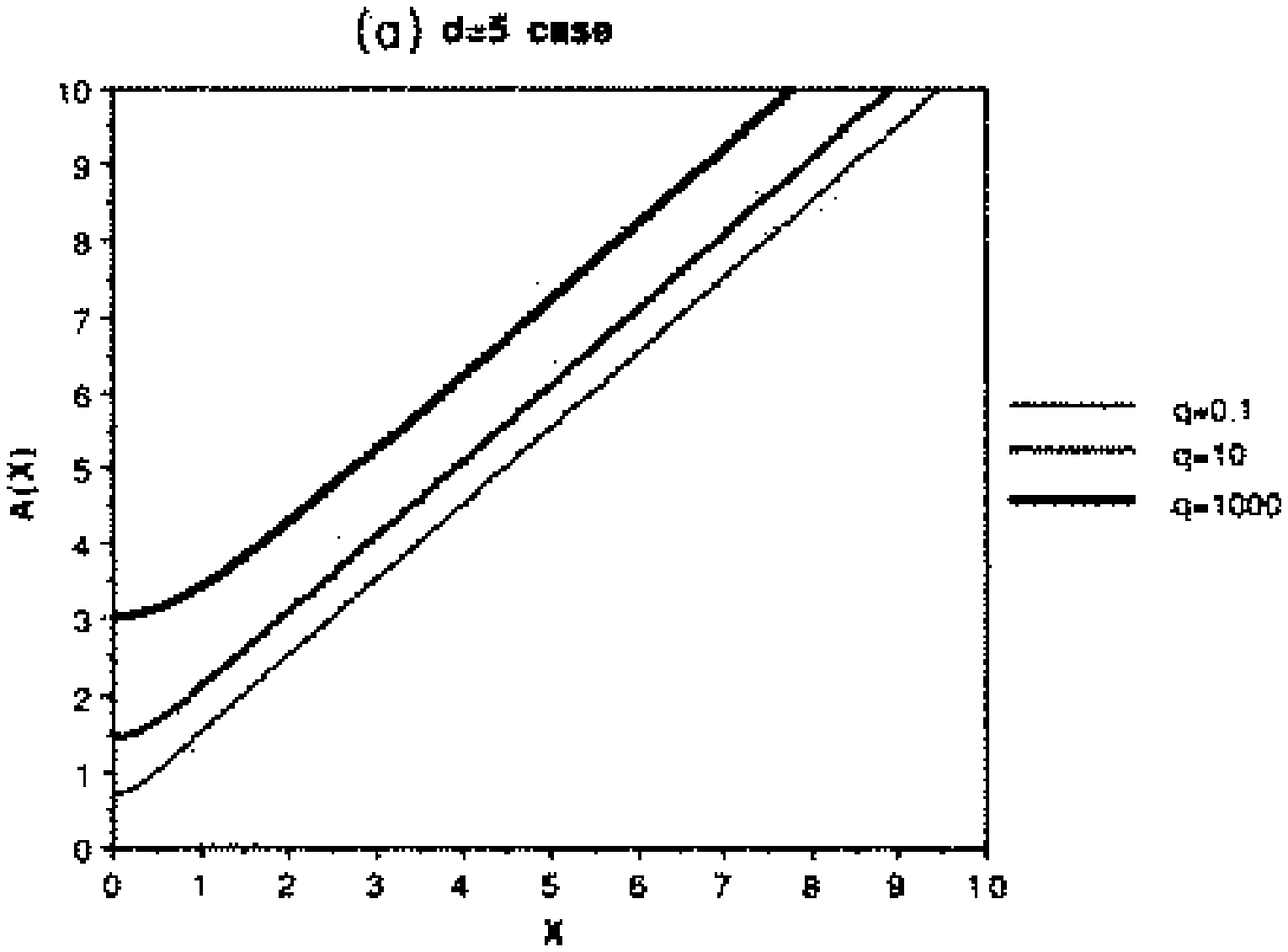}
\includegraphics[width=5cm]{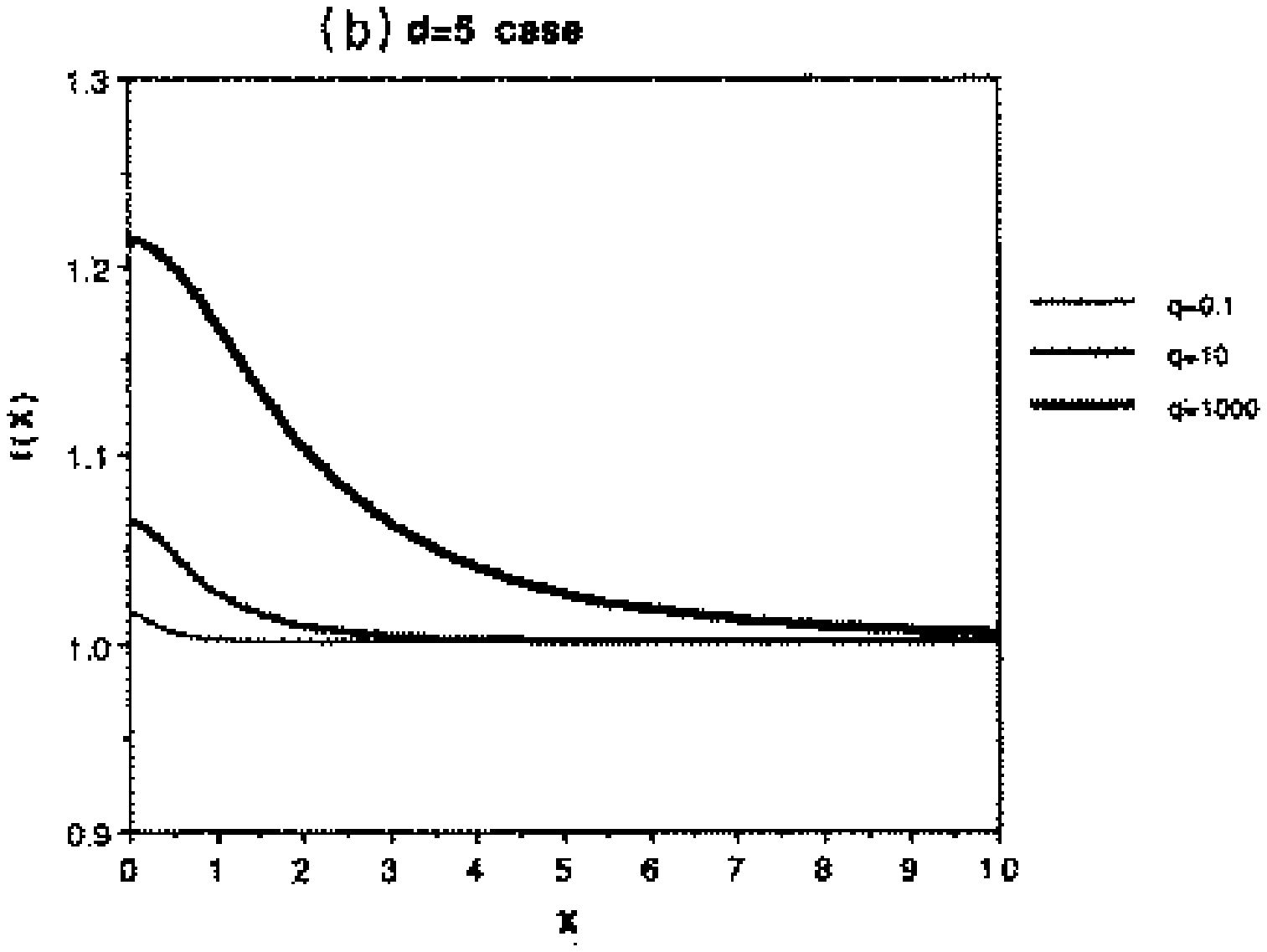}\\
\centering{(a) $A(X)$ as a function of $X$ in the case where the
space dimension is five. The lines correspond to the parameter
$q=1000, 10,$ and $0.1$ in order of boldness. (b) $f(X)$ as a function
of $X$ in the case where the space dimension is five. The
definitions of lines are the same as in (a).}
\label{f2}
\end{center}
\end{figure}
\begin{figure}[ht]
\begin{center}
\includegraphics[width=5cm]{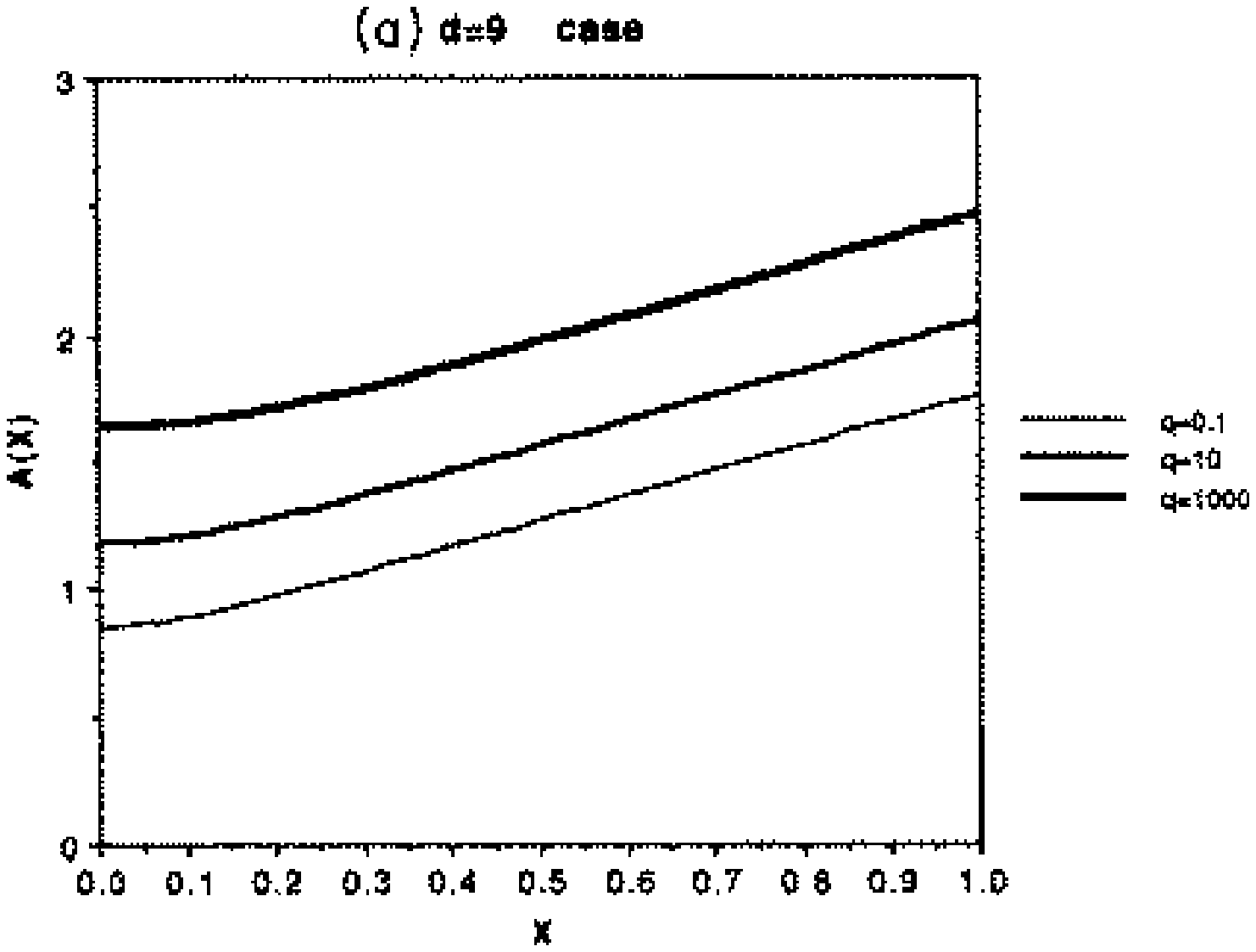}
\includegraphics[width=5cm]{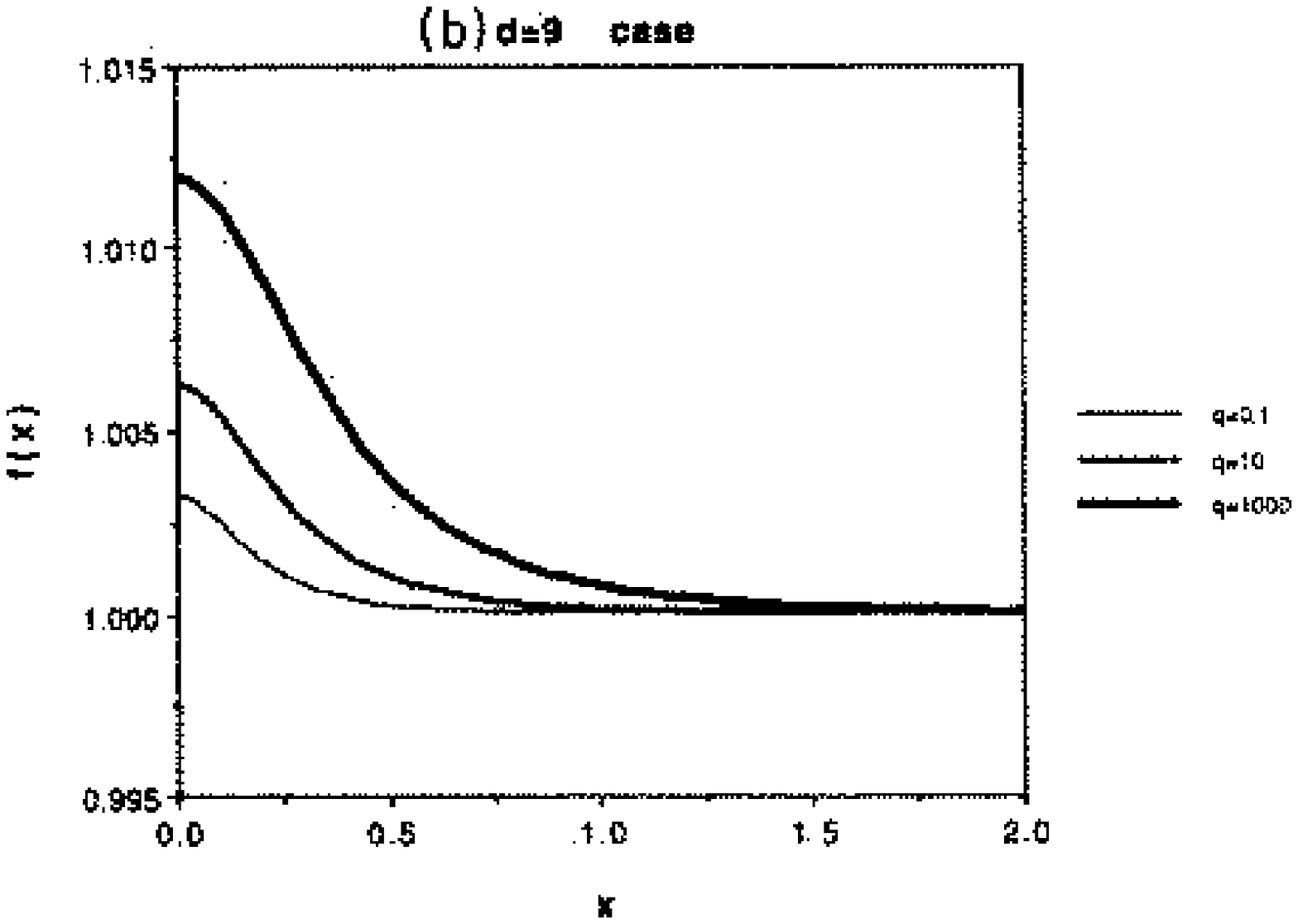}\\
\centering{(a) $A(X)$ as a function of $X$ in the case where the
space dimension is nine. The lines correspond to the parameter
$q=1000, 10,$ and $0.1$ in order of boldness. (b) $f(X)$ as a function
of $X$ in the case where the space dimension is nine. The
definitions of lines are the same as in (a).}
\label{f3}
\end{center}
\end{figure}

Far from the wormhole, i.e., $X\gg 1$, we can ignore
gravity and take the Euclidean space-time to be flat, i.e.,
$A(X)\simeq X$ (Ref.~\cite{10}). But we cannot ignore the effective
potential for the scalar $f$ completely. Here $f$ goes to one,
obeying the effective field equation
\begin{equation}
{f''}+\frac{d-2}{X}{f'}=
\frac{1}{L^2X^2}f(f^2-1)-\frac{2qL^2}{f^3X^{2d -4}}\,.
\label{3.9}
\end{equation}
For large $X$ the first and last terms on the right-hand side
dominate and $f$ is obtained up to the leading order
\begin{equation}
f-1\simeq X^{-\delta}\quad\mbox{with}\quad\delta=
\frac{d-3+\sqrt{(d-3)^2+8/L^2}}{2}\,. 
\label{3.10}
\end{equation}

$f$ in our solutions approaches one according to the negative
power of $X$ at large $X$. It can be contrasted with the
case for the usual scalar wormhole \cite{10}; in the scalar model
with a negative mass term, the value of the scalar field
will decrease exponentially to the stationary value. The
origin of the difference of the behavior at large $X$ is due to
the $A$ dependence of our effective potential.

The case with $d=4$ and $\alpha/\beta=1$ is exceptional, because
of the dominance of the second term on the right-hand
side in (\ref{3.9}). In this case $\delta=2$. In the other cases where
we have performed the numerical calculation, (\ref{3.9}) is always
valid and it gives $\delta=4$ for $d=5$ and $\delta=8$ for $d=9$.

Although $f$ goes to $1$ only as the power of $X$, its contribution
to the action does not diverge. This is because the
effective kinematic term and the potential of the scalar $f$
depends on $A$ in negative power.

Numerical results for $A(0)$ and $f(0)$ as functions of $q$
are shown in Figs.~4(a) and 4(b). We must notice the
dependence on $A$ in (\ref{3.7a}) and (\ref{3.7b}). Owing to the
difference from Eqs.~({1.5}) and ({1.6}) in Ref.~\cite{10},
$f(0)$ increases, rather than decreases, along with the increase of
$q$.

\begin{figure}[ht]
\begin{center}
\includegraphics[width=5cm]{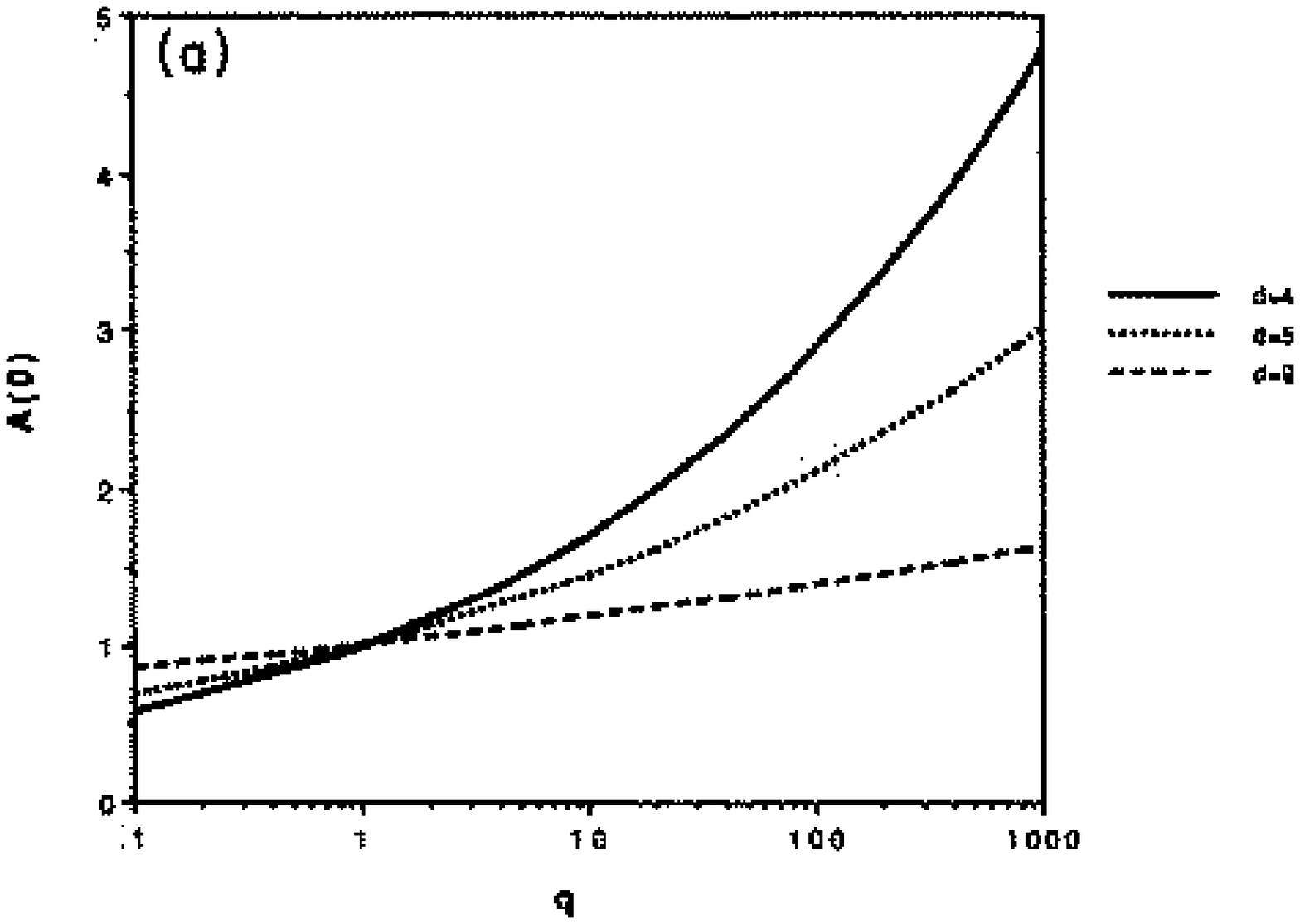}
\includegraphics[width=5cm]{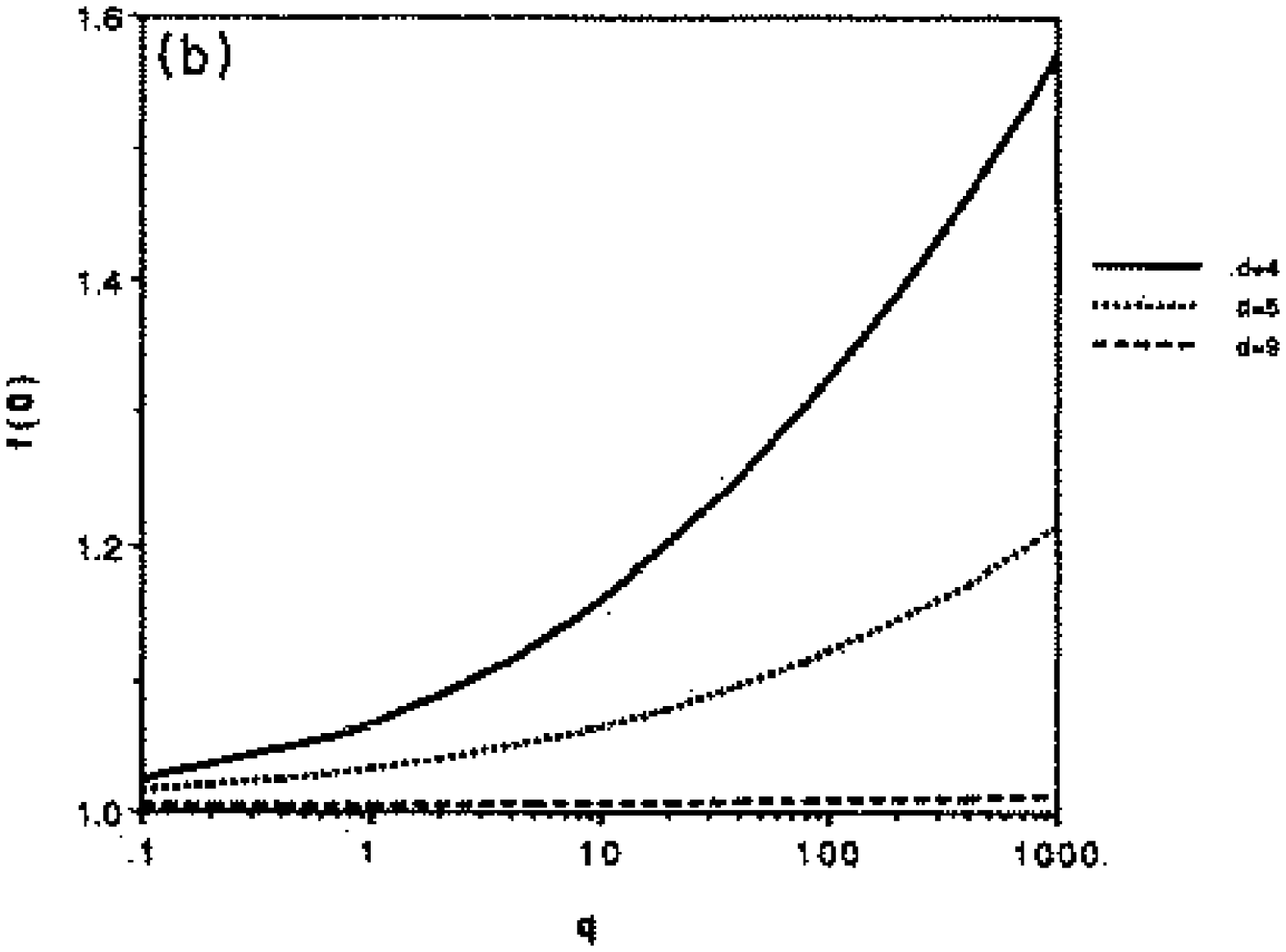}\\
\centering{(a)\hspace{5cm}(b)}
\caption{(a) $A(0)$ as a function of $q$. The solid, dotted, and
dashed lines correspond to the space dimension $d=4, 5,$ and $9$,
respectively. (b) $f(0)$ as a function of $q$. The definitions of lines
are the same as in (a).}
\label{f4}
\end{center}
\end{figure}

These numerical results indicate that the approximation
$f\sim 1$
goes well in the range of $q$, in which we have
performed the computation ($q< 1000$). This means that
we can take
\begin{equation}
A(0)\sim q^{1/(2d-4)}\quad(q< 1000)\, . 
\label{3.11}
\end{equation}

For huge values of $q\gg 1000$, it is expected that
$A(0)\simeq q^{1/(2d-3)}$ and $f(0)\propto q^{2/(2d-3)}$, from the
analysis of the condition derived from (\ref{3.7a}) and (\ref{3.7b}).

$A(0)$ is related to the size of the wormhole ``throat''
and also related to the amount of the action. The action
of the wormhole solution is of order
\begin{equation}
S\sim[A(0)]^{d-1}\propto q^{(d-1)/(2d-4)}\quad
(q < 1000) \,,
\label{3.12}
\end{equation}
and for a huge value of $q$, $S\sim q^{(d-1)/(2d-3)}$.

In Sec.~IV the solutions of Eqs.~(\ref{3.7a}) and (\ref{3.7b}) with
$d=3$ will be examined.

\section{THE FOUR-DIMENSIONAL CASE}
In this section we consider the $d=3$ case, i.e., usual
four-dimensional space-time. In Eqs.~(\ref{3.7a}) and (\ref{3.7b}) in
the preceding section, if we set $d=3$, both the ``centrifugal''
term and the potential term have the same dependence
on the scale factor $A$. Moreover, it is possible to
factorize the power of $A$ and the equation of the motion
for $f$ turns out to be integrable. As a consequence we obtain
\begin{equation}
(Af')'=\frac{1}{2L^2}(f^2-1)^2+\frac{2qL^2}{f^2}-2L^2E\,,
\label{4.1}
\end{equation}
where $E$ is an integration constant and $L^2=\alpha/(2\beta)$.
Using this, we can simplify the equation for $A$:
\begin{equation}
{A'}^2=1-\frac{E}{A^2}\,.
\label{4.2}
\end{equation}
The size of the wormhole throat is given by $E^{1/2}$.

We can solve the equation for $f(X)$ by changing to
``conformal (Euclidean) time'' $dY=dX/A$. We get the
solution in the form of integration: 
\begin{equation}
Y_2-Y_1=\int_{f(Y_1)}^{f(Y_2)}\frac{df}{\sqrt{
(f^2-1)^2/(2L^2)+2qL^2/f^2-2EL^2}}\,.
\label{4.3}
\end{equation}
The shape or behavior of the solutions for $f(Y)$ in quality
can be revealed by inspection of the equation of motion.
We have some different solutions.

If $E<E_0$, where $E_0$ is defined as the equation
\[
(f^2-1)^2/(2L^2)+2qL^2/f^2-2E_0L^2=0\,,
\]
which has a double root, $f(Y)$ increases monotonically
from zero to infinity, or shows the time-reversal behavior,
in a finite-time interval.

If $E> E_0$, there are two types of solutions classified by
the region in which f moves. In one of the cases, $f(Y)$
decreases monotonically from infinity, reaches its
minimum value, and increases to infinity again. In the
other case, $f(Y)$ increases monotonically from zero,
reaches its maximum, and decreases to zero again.

In the case of $E=E_0$, there is a solution with constant
$f$. However, this solution is ``unstable'' in the sense of
classical dynamics.

All the types of solutions, except for the case of
$E =E_0$, are singular since $f$ becomes infinite or zero
within a finite Euclidean time. In other words, these
solutions are singular because ${\rm Tr~}F^2$ diverges.
The consideration of a many-wormhole configuration,
the introduction of Higgs scalars, and the effect of the
terms of higher order in ${\rm Tr~}F^2$ in the action are expected
to cure the singular behavior and bring about ``bounces''
at finite values of the field.

Further investigation will be reported in separate publications.
Here, we shall estimate the interval between
two wormholes or the cutoff scale for a one-wormhole
configuration.

We restrict ourselves on investigating the case with
$E<E_0$ and $f< 1$. The reader will easily perform a similar
analysis in the other case.

The time interval $Y_i$ in which $f$ moves from zero to the
maximum value is obtained by performing the integration
of (\ref{4.3}):
\begin{equation}
Y_i=\frac{\sqrt{2}L}{\sqrt{a-c}}
F\left(\arcsin\left(\frac{b(a-c)}{a(b-c)}\right)^{1/2}, 
\left(\frac{b-c}{a-c}\right)^{1/2}\right)\,,
\label{4.4}
\end{equation}
where $F(\phi, k)$ is the elliptic integral of the first kind \cite{28},
which is defined as
\begin{equation}
F(\phi, k)=\int_0^\phi\frac{d\theta}{\sqrt{1-k^2\sin^2\theta}}\,.
\label{4.5}
\end{equation}

In (\ref{4.4}), $a, b, c$ ($a > b > c$) are the roots of the equation
of the third order, $z^3-2z^2-(4EL^4-1)z+4qL^4=0$. The
cutoff is required to be less than $Y_i$, in order to encounter
the singularity, $f\rightarrow 0$.

In an extreme case, $E \gg q \gg $, $Y_i\simeq (q/2)^{1/2}/EL$, and
the solution is approximately expressed as
\begin{equation}
f^2\simeq (q-8L^2EY^2)/2\,. 
\label{4.6}
\end{equation}

Since the size of the wormhole throat is $E^{1/2}$, the spacetime
manifold bounded by the cutoff looks like a ``ring of
ribbon.'' Therefore it is doubtful that the usual
wormhole dynamics \cite{2,4,5} is applicable in this case.

On the other hand, if $E$ approaches $E_0$, $Y_i$ grows
logarithmically. When $q$ is much smaller than one, we get
\begin{equation}
Y_i\simeq\frac{L}{2\sqrt{2}}\ln\frac{1}{4(E-E_0)L^4}\,,
\label{4.7}
\end{equation}
with $E_0\simeq q$, whereas $q$ is much larger than one, we have
\begin{equation}
Y_i\simeq\frac{1}{2\sqrt{2}(12E_0)^{1/4}}
\ln\frac{3E_0}{E-E_0}\,,
\label{4.8}
\end{equation}
with $E_0\simeq\frac{3}{4}(2qL^4)^{2/3}/L^4$. In these cases we can
take a large cutoff scale.

We can introduce the cosmological constant instead
using the cutoff. Then the equation for $A$ is modified as
\begin{equation}
{A'}^2=1-\frac{E}{A^2}-\frac{\Lambda}{3}A^2\,,
\label{4.9}
\end{equation}
where $\Lambda$ is the cosmological constant. We consider that
a wormhole attaches to a de Sitter universe of which the
radius is approximately $\simeq(3/\Lambda)^{1/2}$. Periodic wormhole
solutions with the cosmological constants are considered
by authors of Refs.~\cite{14} and \cite{18}. One period is likely to be
less than or equal to $Y_i$, the period of the matter
configuration.

We assume that the effect of a higher-derivative term
or other effects work in order to avoid the singular behavior
inside the de Sitter universe. The present solutions
may be cut at about the maximum volume and sewn to
the solution of the equation in which higher-derivative
terms or other dynamics dominate.

We estimate the interval between the universes. The
scale factor $A$ grows from minimum to maximum value
in the conformal time interval $Y_u$. When $A$ is small, we
have \cite{16,17}
\begin{equation}
Y_u\simeq\ln\frac{\sqrt{3}}{4\sqrt{\Lambda E}}\,.
\label{4.10}
\end{equation}
Therefore the plausible size of the cosmological constants is
\begin{equation}
\Lambda>
\frac{3}{16E}\left[ \frac{1}{4(E-E_0)L^4}\right]^{L/2\sqrt{2}}\,,
\label{4.11}
\end{equation}
with $E_0\simeq q$, whereas $q$ is much larger than one, we have
\begin{equation}
\Lambda>
\frac{3}{16E}\left[
\frac{3E_0}{E-E_0}\right]^{L/2\sqrt{2}(12E_0)^{1/4}}\,.
\label{4.12}
\end{equation}

We have estimated the interval between two
wormholes, assuming the existence of nonsingular solutions
for the Yang--Mills field when the modification of
the dynamics is expected. It is necessary to study multiwormhole
effects and/or higher-derivative modifications
near the singular behavior of our solution in future work.

In the next section we will turn to the case of $d=2$, i.e.,
the three-dimensional case, and look for nonsingular
wormhole solutions.

\section{THREE-DIMENSIONAL WORMHOLE
AND CHERN-SIMONS TERM}
In three space-time dimensions, there is no wormhole
solution of the type which we treated in Sec.~III in a pure
Yang--Mills system. If we add $U(1)$ gauge fields to the action
we can obtain the wormhole solutions which are
kept from collapse by the magnetic charge \cite{16,21}.

Here, we do not require the introduction of other
fields. Instead, we consider the action which includes the
Chern--Simons term \cite{29} as well as the conventional Yang--Mills
term.

We consider $SU(2)$ Yang--Mills coupled to gravity. The
action is
\begin{eqnarray}
S&=&\int d^3x\left\{\sqrt{-g}\left[-\frac{1}{2\kappa^2}R+
\frac{1}{4e^2}{\rm Tr~}F^{MN}
F_{MN}\right]\right.\nonumber \\
& &\qquad\left.+\frac{\mu_E}{2}\epsilon^{MNL}
(A_MF_{NL}+\cdots)\right\} +(\mbox{surface terms})\,,
\label{5.1}
\end{eqnarray}
where the ellipsis denotes the term which consists of a triple
product of gauge fields.

Because the Chern--Simons term in three dimensions is
topological, i.e., it does not include the metric, the Einstein
equation is the same as those of pure Yang--Mills
theory. The Yang--Mills equations, however, change their
form, and also the integration in terms of the ``cyclic''
variable is modified. The coefficient $\mu_E$ of the Chern-Simons term
is pure imaginary if the time direction is already taken as Euclidean.

We again use the spherical ansatz on the metric, i.e.,
\begin{equation}
ds^2=d\tau^2 +a^2(\tau)(d\theta^2+\sin^2 d\phi^2) \,.
\label{5.2}
\end{equation}
The symmetric reduction of the $SU(2)$ gauge field on $S^2$
produces a $U(1)$-symmetric potential with a wine-bottle
(or Mexican-hat) shape. This is because $S^2=SO(3)/SO(2)$
and $SU(2)\simeq SO(3)$. Thus this is a ``minimal'' example for
the reduction of gauge symmetry.

To express the mapping explicitly we write the gauge
fields as
\begin{eqnarray}
A_\theta&=&\frac{1}{2}\left(
\begin{array}{cc}
0 & -i\Phi e^{-i\phi}\\
i\Phi^*e^{i\phi} & 0
\end{array}
\right)\,,\label{5.2a}\\
A_\phi&=&-\frac{1}{2}\left(
\begin{array}{cc}
0 & \Phi e^{-i\phi}\\
\Phi^*e^{i\phi} & 0
\end{array}
\right)\sin\theta+\frac{1}{2}\left(
\begin{array}{cc}
1-\cos\theta & 0\\
0 & -(1-\cos\theta)
\end{array}
\right)\,,
\label{5.2b}
\end{eqnarray}
where $\theta$ and $\phi$ are the polar and the azimuthal angles of
the sphere. Note that here we use a coordinate basis associated
with the metric and not an orthonormal one.
This mapping leads to the effective Lagrangian of the
gauge field
\begin{eqnarray}
& &\frac{1}{4e^2}{\rm Tr~}F^{MN} F_{MN}+
\frac{\mu_E}{2\sqrt{+g}}\epsilon^{MNL}(A_MF_{NL}+\cdots)\nonumber \\
& &=2\frac{|\dot{\Phi}|^2}{a^2}+\frac{(|\Phi|^2-1)^2}{a^4}+
\frac{\mu}{2a^2}i(\Phi^*\dot{\Phi}-\phi\dot{\Phi}^*)\,,
\label{5.3}
\end{eqnarray}
where $\mu=i\mu_E$.

Gauge invariance of the quantum theory, which is
defined by the path integral, requires quantization of the
coupling constant $\mu$. We do not take this fact into account,
since we treat only classical equations.

If we write $\Phi$ as $\Phi=fe^{i\psi}$, then we can integrate the
equation on $\psi$ and we obtain
\begin{equation}
\frac{1}{e^2} f^2\dot{\psi}=i\mu f^2+in\,, 
\label{5.4}
\end{equation}
where $n$ is an integration constant.

We take a unit basis of length so that $(\kappa^2/e^2)|n/\mu|$
equals $1$. Moreover we define the parameter set as
$\epsilon=|n/\mu|$ and $m^2=\mu^2 e^4$. To emphasize the change of the
unit basis we write the scale factor as $A$. The scalar variable
is rescaled as $F^2\equiv f^2/\epsilon$.

The field equations are written by using (\ref{5.4}) as
\begin{eqnarray}
\ddot{A}&=&\frac{1}{4\epsilon A^3}(\epsilon F^2-1)^2\,,
\label{5.5a}\\
\ddot{F}&=&m^2 F\left(1-\frac{1}{F^4}\right)+\frac{1}{A^2}
F(\epsilon F^2-1)\,.
\label{5.5b}
\end{eqnarray}
We consider the case with $\epsilon<1$ only.

The behavior of the possible solution can be seen from
(\ref{5.5a}) and (\ref{5.5b}). As a wormhole solution we require that
$A$ diverges linearly when $\tau$ goes to infinity. Then the
value $F$ is attracted to $1$ at large $\tau$. Since the original
variable $f$ goes to $\sqrt{\epsilon}$, the gauge symmetry is broken at
the asymptotic region.

\begin{figure}[ht]
\begin{center}
\includegraphics[width=5cm]{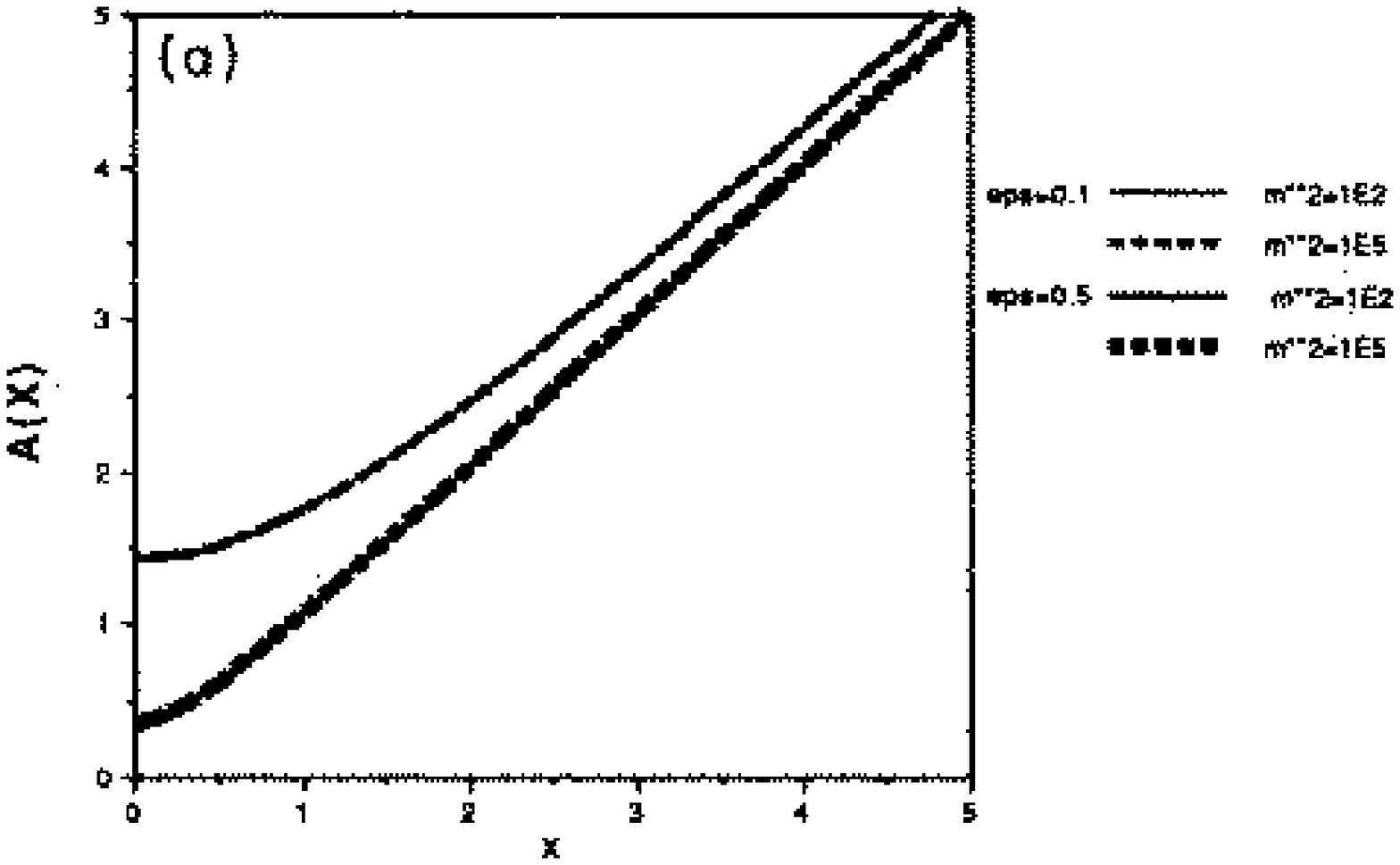}
\includegraphics[width=5cm]{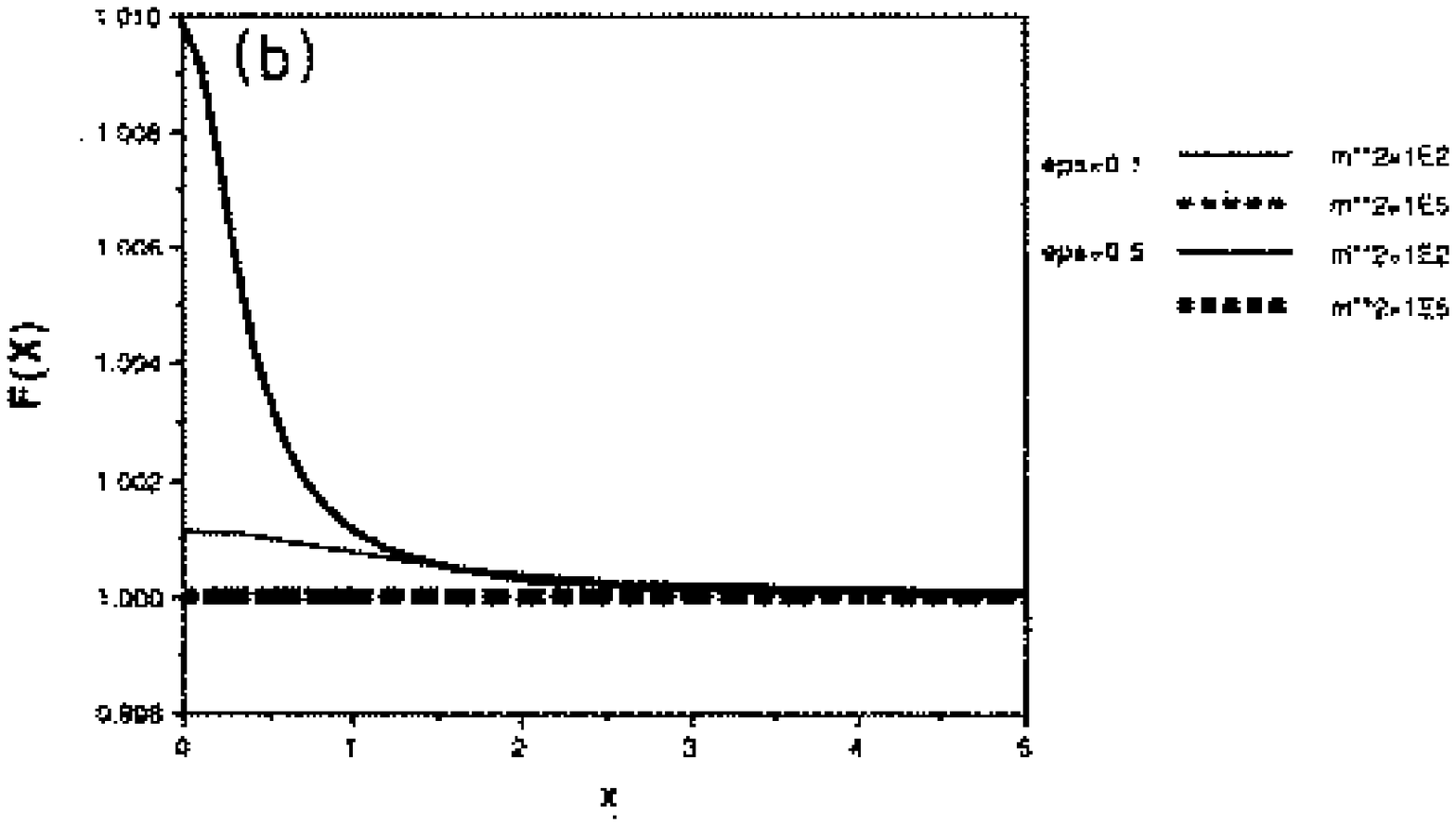}\\
\centering{(a)\hspace{5cm}(b)}
\caption{(a) $A(X)$ as a function of $X$ in the case where the
space dimension is two. The solid and dashed lines correspond
to the parameter $m^2=10^2$ and $10^5$. At the same time, the thin
and thick lines correspond to the parameter $\epsilon=0.1$ and $0.5$.
(b) $F(X)$ as a function of $X$ in the case where the space dimension is
two. The definitions of lines are the same as in (a). The dashed
lines overlap one another.}
\label{f5}
\end{center}
\end{figure}

The result of numerical calculations is shown in Figs.~5(a) and 5(b).
If e increases, $A(0)$ decreases. On the other hand, $A(0)$ is almost
independent of the value of $m^2$. No wormhole solution exists if m is
less than a certain critical value. Because $F(0)$ grows when m
decreases and $F$ cannot grow beyond $1/\sqrt{\epsilon}$, the critical
value exists. Therefore the value depends on $\epsilon$. The details
will be published elsewhere.

The action of these wormholes diverges but the divergence
is so-called ``infrared''; namely, an arbitrary finite
cutoff scale (or cosmological constant) makes the action
\cite{10,16,17}. 

The results obtained in this section are rather
``academic'' ones, but the model we considered is very
useful to investigate the connection of matter and gravity
in the three-dimensional system, which has attracted
much attention recently.

\section{CONCLUSIONS}
We have examined wormhole solutions in non-Abelian
gauge theories with a sufficiently large symmetry group.
We find that wormhole solutions which are very similar
to the scalar wormholes exist in the case where the spatial
dimension is larger than three. In these solutions, the
gauge symmetry is unbroken in the asymptotic region
and the Euclidean action of the solutions is finite. In the
case that the spatial dimension equals three, we obtain
the solution with the infinite action. The divergence of
the action is stronger than the usual divergence \cite{10}, because
the Lagrangian density of the wormhole solutions
itself diverges at finite Euclidean time. This fact suggests
that we must consider the ``many-body problem'' of
wormholes seriously in the Yang--Mills system of four dimensional
space-time.

The known solutions of Refs.~\cite{16} and \cite{17} in the $SU(2)$
Yang--Mills system have an ordinary ``infrared divergence''
and a finite Lagrangian density. We consider that
our model is more generic than those of Refs.~\cite{16} and \cite{17};
their model with one effective real scalar is applied to the
case with a minimal size of gauge symmetry such as
$SU(2)$ on $S^3$, while our model with an effective complex
scalar can be applied to the case with the arbitrary large
gauge group. That is because of the existence of a term
such as ``centrifugal potential'' in the effective theory
after the reduction of gauge fields of sufficiently large
symmetry.

To avoid the singular behavior we may consider
higher-order terms in $F_{MN}$. Note that the finite action
can be obtained in the case where the spatial dimension
slightly deviates such that $d =4+\epsilon~(\epsilon>0)$. Running of
the gauge coupling constant may change the behavior of
the solutions, and the dilaton coupled to the Yang--Mills
fields also affects the behavior of the solutions. These are
interesting future problems to study.

\section*{ACKNOWLEDGMENTS}
The authors would like to thank A.~Nakamula for
some useful comments. This work was supported in part
by the Grant-in-Aid for Encouragement of Young Scientists
from the Ministry of Education, Science and Culture
(Grant No.~63790150). One of the authors (K.S.) is grateful
to the Japan Society for the Promotion of Science for
financial support. He also thanks Iwanami F\=ujukai for
financial aid.

\section*{APPENDIX}
In this appendix we will show an example of dimensional
reduction of non-Abelian gauge fields on $S^d$.
Please note that this example is a simple one to express
explicit representation, but this reduction may not be a
minimal example to obtain a $U(1)$-symmetric effective potential.

We consider the $SU(N)$ Yang--Mills gauge fields in the
space of $S^d$. Here $N$ must be greater than or equal to
$d+1$.

We express the gauge potential as an $N \times N$ matrixvalued
one-form $\mathbf{A}$. The generators of $SO(d +1)$ are divided
into the generators of $SO(d)$ and those of the complements
of $SO(d)$ in $SO(d+1)$. According to this we assume
that $\mathbf{A}$ can be decomposed as
\begin{equation}
\mathbf{A}=\mathbf{A}_Ae^A=\mathbf{A}_ie^i+\mathbf{A}_ae^a\,,
\label{A1}
\end{equation}
where $e^A$ are one-forms which satisfy
\begin{equation}
de^A=\frac{1}{2} f^{ABC}e^B\wedge e^C\,,
\label{A2}
\end{equation}
where $f^{ABC}$ is the structure constant of $SO(d+1)$.

Among them, $e^i$ ($i=1,\dots, d$) transform homogeneously
with the $SO(d)$ rotation, i.e.,
\begin{equation}
de^i=f^{iaj}e^a\wedge e^j\,. 
\label{A3}
\end{equation}

On the other hand, $e^a$, $a=1, \dots, d(d-1)/2$, obey the
Mauer--Cartan equations
\begin{equation}
de^a=\frac{1}{2}f^{abc}e^b\wedge e^c+\frac{1}{2}f^{aij}e^i\wedge
e^j\,.
\label{A4}
\end{equation}
Note that $f^{abc}$ is the structure constant of the $SO(d)$ subgroup.

The field strength is defined as
\begin{equation}
F=d\mathbf{A}+\frac{1}{2}i\mathbf{A}\wedge\mathbf{A}\,.
\label{A5}
\end{equation}
We require the background configuration of the gauge
field. We assume that $\mathbf{A}_a$ satisfies
\begin{equation}
[\mathbf{A}_a, \mathbf{A}_b]=if^{abc}\mathbf{A}_c\,, 
\label{A6}
\end{equation}
where $f^{abc}$ is the structure constant of $SO(d)$. These
configurations can be represented by d Xd submatrices.
Those are just the generators with the appropriate normalization.
Further the standard representation of the
$SO(d)$ generators in the $d\times d$ matrix form is assumed.
Then we set
\begin{equation}
(\mathbf{A}_i)_{\alpha\beta}=i(\delta_{\alpha,d+1}\delta_{\beta,i}
\Phi
-\delta_{\alpha,i}\delta_{\beta,d+1}\Phi^*) 
\label{A7}
\end{equation}
where $\alpha$ and $\beta$ denote the component of the matrix. Here
we mainly considered the $SO(d+1)$ subgroup of $SU(N)$
and we adopted an additional ``phase'' to make the complex
variable $\Phi(\tau)$.

Taking above ``ansatz'' with the background geometry
$R\times S^d$, we obtain
\begin{equation}
{\rm Tr~} F_{AB} F^{AB}=4d|\dot{\Phi}|^2/a^2
+2d(d-1)(1-|\Phi|^2)^2/a^4\,,
\label{A8}
\end{equation}
where $a$ is the radius of $S^d$ and the overdot denotes the
derivative with respect to the Euclidean time $\tau$. Equation
(\ref{A8}) corresponds to Eq.~(\ref{2.2}) in the text with
$\alpha=\beta=1$.


\end{document}